\RequirePackage{fix-cm}
\documentclass[twocolumn,natbib]{svjour3}          
\smartqed  
\usepackage{graphicx}
\usepackage{amsfonts,amssymb,amsmath,times}
\usepackage{doi}
\journalname{Climate Dynamics}
\begin{document}

\title{How complex climate networks complement eigen techniques for the statistical analysis of climatological data
}


\author{Jonathan F. Donges \and Irina Petrova \and Alexander Loew \and Norbert Marwan \and  J\"urgen Kurths}

\authorrunning{Donges, Petrova et al.} 

\institute{Jonathan F. Donges \and Norbert Marwan \and J\"urgen Kurths
  \at Potsdam Institute for Climate Impact Research, P.O. Box 60\,12\,03, 14412 Potsdam, Germany\\
  \email{donges@pik-potsdam.de}
  \and Jonathan F. Donges
  \at Stockholm Resilience Center, Stockholm University, Kr\"aftriket 2B, 114\,19 Stockholm, Sweden
  \and Irina Petrova
  \at Max-Planck-Institute for Meteorology, KlimaCampus, 20146 Hamburg, Germany
  \and Alexander Loew
  \at Department of Geography, University of Munich (LMU), Luisenstr.~37, 80333 Munich, Germany
  \and J\"urgen Kurths
  \at Department of Physics, Humboldt University, Newtonstr.~15, 12489 Berlin, Germany
  \at Institute for Complex Systems and Mathematical Biology, University of Aberdeen, Aberdeen AB243UE, United Kingdom
  \at Department of Control Theory, Nizhny Novgorod State University, 603950 Nizhny Novgorod, Russia}

\date{Received: date / Accepted: date}

\maketitle

\begin{abstract}
Eigen techniques such as empirical orthogonal function (EOF) or coupled pattern (CP) / maximum covariance analysis have been frequently used for detecting patterns in multivariate climatological data sets. Recently, statistical methods originating from the theory of complex networks have been employed for the very same purpose of spatio-temporal analysis. This climate network (CN) analysis is usually based on the same set of similarity matrices as is used in classical EOF or CP analysis, \textit{e.g.}, the correlation matrix of a single climatological field or the cross-correlation matrix between two distinct climatological fields. In this study, formal relationships as well as conceptual differences between both eigen and network approaches are derived and illustrated using global precipitation, evaporation and surface air temperature data sets. These results allow us to pinpoint that CN analysis can complement classical eigen techniques and provides additional information on the higher-order structure of statistical interrelationships in climatological data. Hence, CNs are a valuable supplement to the statistical toolbox of the climatologist, particularly for making sense out of very large data sets such as those generated by satellite observations and climate model intercomparison exercises.
\keywords{climate networks \and empirical orthogonal functions \and coupled patterns \and maximum covariance analysis \and climate data analysis}
\end{abstract}

\section{Introduction}

Climatologists have long been interested in studying correlations between climatological variables for gaining an understanding of the Earth's climate system's large-scale dynamics~\citep{Katz2002}. Pioneering work in this field was done by Sir Gilbert T. Walker in the beginning of the 20th century while attempting to find precursory patterns for Indian monsoon events using statistical methods~\citep{Walker1910}, which culminated in the discovery of the tropical Walker circulation and the Pacific Southern Oscillation (a part of the El Ni{\~n}o-Southern Oscillation known as ENSO). Later, new measurement devices as well as the rapid increase in available computing power allowed to investigate statistical interdependency structures of global or regional climatological fields $\mathbf{x}(t) = \{x_i(t)\}_{i=1}^N$ such as surface air temperature, pressure, or geopotential height~\citep{Fukuoka1951,Lorenz1956} (here, $i$ is a spatial index, \textit{e.g.}, labeling $N$ meteorological measurement stations or grid points in an aggregated data set, and $t$ denotes time).

Nowadays, techniques of eigenanalysis such as empirical orthogonal functions (EOFs)~\citep{Kutzbach1967,Wallace1981,Hannachi2007} and coupled patterns (CPs)~\citep{Bretherton1992} are standard tools for finding spatial as well as temporal patterns in climatological data~\citep{vonStorch2003}. Their applications range from statistical predictions~\citep{Lorenz1956,Brunet1996,Repelli2004}, over the definition of climate indices~\citep{Power1999,Leroy2008} to evaluating the performance of climate model simulation runs~\citep{Handorf2009,Handorf2012}. While numerous linear and nonlinear extensions have been proposed~\citep{Ghil1991,Ghil2002}, e.g., rotated or simplified EOFs~\citep{Hannachi2007} and other methods of dimensionality reduction such as neural network-based nonlinear principal component analysis (PCA) \citep{Hsieh2004} or isometric feature mapping (ISO\-MAP) \citep{Tenenbaum2000,Gamez2004}, classical EOF and CP analysis have remained among the most popular statistical techniques applied in climatology so far.

In the last decade, complex network theory has been introduced as a powerful framework for extracting information from large volumes of high-dimensional data~\citep{Newman2003,Boccaletti2006,Newmanbook2010,Cohenbook2010} such as those generated by neurophysiological or biochemical measurements, quantitative social science as well as climatological observations and modeling campaigns. While EOFs, CPs, and related methods effectively rely on a dimensionality reduction, network techniques allow to study the full complexity of the statistical interdependency structure within a multivariate data set. In these climate networks (CNs), which were first introduced by \citet{Tsonis2004,Tsonis2006}, nodes correspond to time series of climate variability at grid points or observational stations and links indicate a relevant statistical association between two such time series. For quantifying statistical associations, linear covariance or Pearson correlation can be used analogously to EOF and CP analysis~\citep{Tsonis2004,Tsonis2008a,Yamasaki2008}, but nonlinear measures such as mutual information~\citep{Donges2009b,Donges2009a,Barreiro2011} or transfer entropy~\citep{Runge2012} may be employed as well with care~\citep{Hlinka2014}. Among other applications, CNs have been used to uncover global impacts of El Ni\~no events~\citep{Tsonis2008a,Yamasaki2008,Gozolchiani2011,Martin2013,Radebach2013}, trace the flow of energy and matter in the surface air temperature field~\citep{Donges2009b}, unravel the complex dynamics of the Indian summer monsoon~\citep{Malik2012,Stolbova2014}, detect community structure enabling statistical prediction of climate indices~\citep{Tsonis2010,Steinhaeuser2011b,Steinhaeuser2011} as well as intercomparisons between climate models and observations~\citep{Steinhaeuser2013,Feldhoff2014}, and study large-scale circulation patterns and prominent modes of variability in the atmosphere~\citep{Tsonis2008b,Donges2011a,EbertUphoff2012a,EbertUphoff2012b}. Furthermore, CN analysis has recently been employed to improve forecasting of El Ni\~no episodes~\citep{Ludescher2013,Ludescher2014}, predict extreme precipitation events over South America~\citep{Boers2014natcom} and to derive early warning indicators for the collapse of the Atlantic meridional overturning circulation~\citep{Mheen2013}. Extending upon the majority of studies focussing on recent climate variability, the CN approach has also been applied to study late Holocene Asian summer monsoon dynamics based on data from paleoclimate archives~\citep{Rehfeld2013}

The main aim of this contribution is to put the recent CN approach into context with standard eigenanalysis, since both classes of methods are often based on the same set of statistical similarity matrices. We briefly review both classes of techniques to establish a common notation. Formal relationships are then derived between empirical orthogonal functions or coupled patterns and frequently used CN measures such as degree or cross-degree, respectively. These relationships are illustrated empirically using global satellite observations of precipitation and evaporation fields as well as surface air temperature reanalysis data. We furthermore illustrate and argue in which settings higher-order CN measures such as betweenness may contain information complementing classical eigenanalysis. For example, betweenness can be interpreted as approximating the flow of energy and matter within a climatological field and is particularly useful for identifying bottlenecks that may be particularly vulnerable to perturbations such as volcanic eruptions or anthropogenic influences~\citep{Donges2009b,Donges2011a,boers2013complex,Molkenthin2014a}. Hence, by transferring insights and tools from complex network theory and complexity science to climate research, CNs meet the need for novel techniques of climate data analysis facing quickly increasing data volumes generated by growing observational networks and model intercomparison exercises like the coupled model intercomparison project (CMIP)~\citep{Meehl2005,Taylor2012}.

This article is structured as follows: After describing the data to be analyzed (Section~\ref{sec:data}), we introduce eigen (Section~\ref{sec:eigen}) and network (Section~\ref{sec:network}) techniques for the statistical analysis of climatological data. Relationships between both approaches are formally derived and empirically demonstrated using observational climate data in Section~\ref{sec:relationships}. This leads us to pinpoint the added value of CN analysis (Section~\ref{sec:benefits}), before concluding in Section~\ref{sec:conclusions}.

\section{Data} \label{sec:data}

Imperfect retrieval algorithms and data merging of atmospheric fields that are involved in the generation of reanalysis data sets may cause uncertainties and lower quality of the final product of data analysis. In order to obtain consistent and representative precipitation and evaporation fields, in this study, the fully satellite-based HOAPS-3 (Hamburg Ocean Atmosphere Parameters and Fluxes from Satellite Data, http://www.hoaps.org, \citet{Andersson2010,Andersson2011}) and combined HOAPS-3/ GPCC (Global Precipitation Climatology Center, http://www.gp\-cc.dwd.de, \citet{Andersson2010b}) data sets are used. Regardless of the improved retrieval algorithms and high quality output product, the uniqueness of the HOAPS data set consists in utilization of only one satellite data set for retrieval of both, evaporation, and precipitation parameters. Originally available at the resolution of 0.5 degrees in latitude and longitude, monthly mean precipitation ($\mathbf{x}$(t)) and evaporation ($\mathbf{y}$(t)) anomaly fields ($1992$--$2005$) were resampled to T63 resolution ($\approx 1.8$ degrees) to reduce computational costs. Furthermore, areas with sea-ice coverage were excluded from the set of raw time series. This results in $N_P=13,834$ and $N_E=7,986$ grid points (or network nodes) and $M=168$ samples for each time series for the global precipitation and evaporation data sets, respectively. The smaller number of nodes in the evaporation field arises because the data are only available over the oceans, but not over land. We use the full global data sets for comparing univariate techniques of climate data analysis, but for clarity restrict ourselves to the North Atlantic Ocean region for the multivariate methods.

Additionally, to put our work into context with earlier work on CN analysis~\citep{Tsonis2008a,Yamasaki2008,Donges2009b,Steinhaeuser2011}, we study global monthly averaged surface air temperature (SAT) field data covering the years 01/1948--12/2007 taken from the reanalysis I project provided by the  National Center for Environmental Prediction / National Center for Atmospheric Research (NCEP/NCAR, \citet{Kistler2001}). This data set consists of $N_T= 10,224$ grid points (network nodes) and $M=720$ samples for each time series.

\section{Eigenanalysis} \label{sec:eigen}

This section serves to introduce the mathematics of eigenanalysis necessary for the deductions made below. Spe\-cifically, standard EOF analysis of single climatological fields (e.g., the precipitation field) as well as coupled patterns based on a singular value decomposition of the cross-correlation matrix (also termed \emph{maximum covariance analysis} (MCA) in \citet{vonStorch2003}) for studying statistical relationships between two climatological fields (e.g., the precipitation and evaporation fields) are discussed. Of all the variants of eigenanalysis~\citep{Hannachi2007}, these two approaches appear to be the most frequently used and are also most closely related to CN and coupled CN analysis, respectively, as will be elaborated on in Section~\ref{sec:relationships}. For further details, the reader is referred to \citet{Bretherton1992,vonStorch2003} or \citet{Hannachi2007}.

Note, that for consistency with the CN literature (see Section~\ref{sec:network}), we define EOFs (CPs) based on the correlation (cross-correlation) instead of the covariance (cross-covariance) matrix. The results and conclusions presented in Sections~\ref{sec:relationships} and \ref{sec:benefits} would not change qualitatively if the covariance (cross-covariance) matrix would be used for both eigenanalysis and CN construction.

\subsection{Empirical orthogonal function analysis}

\begin{figure}[tb]
\begin{center}
\includegraphics[width=\columnwidth]{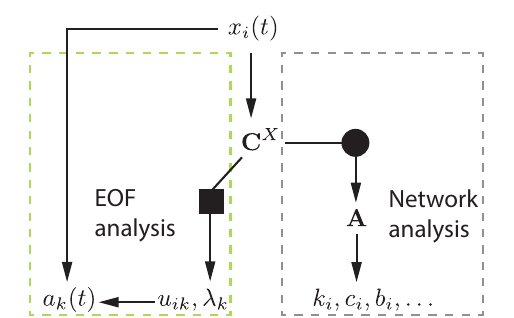}
\caption{A schematic outline of the relationship between univariate EOF and climate network analysis in the spirit of the diagrams in \citet{Bretherton1992}. The eigen decomposition (PCA) operation is represented by the square, the thresholding operation by the disc. All vectors are written in component form.}
\label{fig:univariate_scheme}
\end{center}
\end{figure}

Given a set of normalized time series $\mathbf{x}(t) = \{x_i(t)\}_{i=1}^N$ with zero mean and unit standard deviation, the \emph{correlation matrix} $\mathbf{C}^X=\{C_{ij}^X\}_{ij}$ is defined by

\begin{equation}
C_{ij}^X = \frac{1}{M} \sum_{t=1}^M x_i(t) x_j(t), \label{eq:correlation_matrix}
\end{equation}
where $M$ is the length (number of samples) of each time series.

The aim of EOF analysis (also termed principal component analysis in the statistical literature~\citep{Preisendorfer1988}) is a dimensional reduction achieved by decomposing the data into linearly independent linear combinations of the different variables that explain maximum variance~\citep{Hannachi2007}. The EOFs $\mathbf{u}_k$ are obtained as solutions of the eigenvalue problem

\begin{equation}
\mathbf{C}^X \mathbf{u}_k = \lambda_k \mathbf{u}_k.
\end{equation}
The $k$--th EOF $\mathbf{u}_k$ is the eigenvector corresponding to the $k$--th largest eigenvalue $\lambda_k$, where $u_{ik}$ denotes the $i$--th component of the $k$--th EOF (Fig.~\ref{fig:univariate_scheme}). The EOFs are sorted according to the ordering of their associated non-negative eigenvalues $\lambda_k$ such that $\lambda_1 \geq \lambda_2 \geq \dots \geq \lambda_R$ ($R$ is the rank of $\mathbf{C}^X$). Hence, $\mathbf{u}_1$ associated with the largest eigenvalue $\lambda_1$ is called the \emph{leading EOF} of the underlying data set and represents the one-dimensional projection of the data with the largest possible variance.

\begin{figure}[tb]
\begin{center}
\includegraphics[width=\columnwidth]{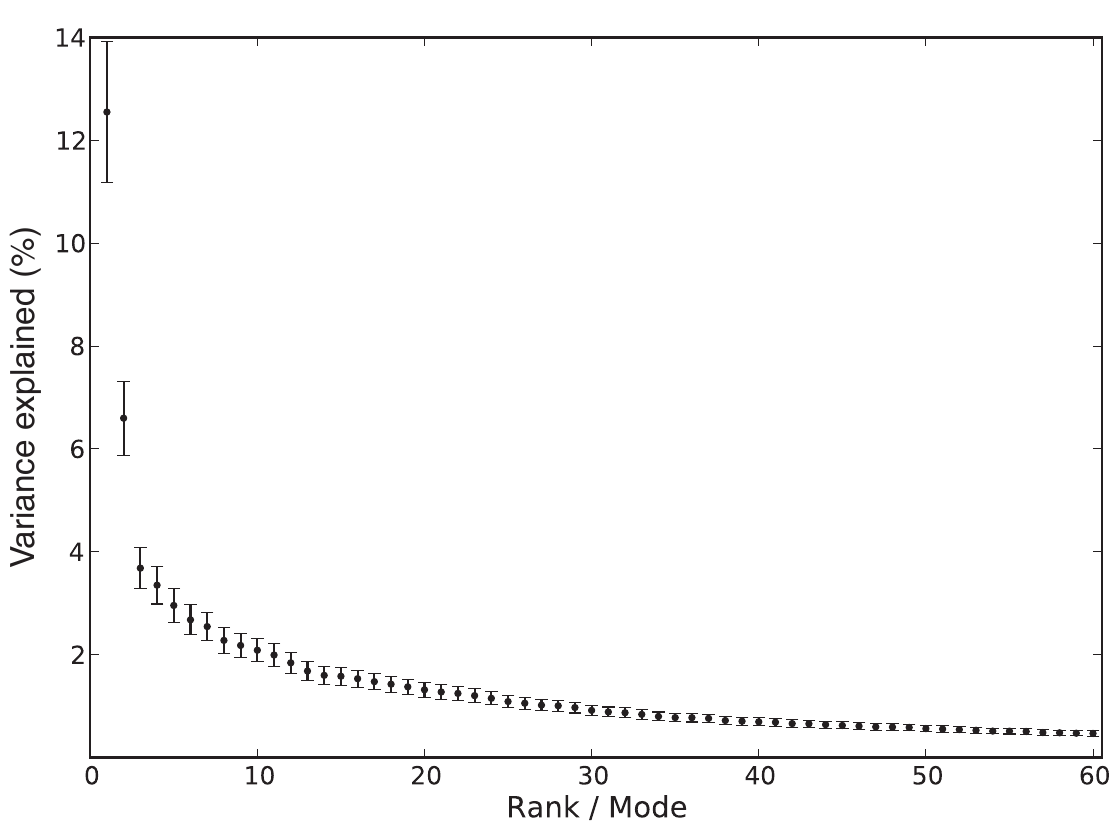}
\caption{Percentage of variance $\lambda_k / \sum_{l=1}^R \lambda_l$ explained by EOFs $\mathbf{u}_k$ for the HOAPS-3 / GPCC precipitation data set. Error bars were estimated using North's rule of thumb~\citep{North1982}.}
\label{fig:eigen_spectrum}
\end{center}
\end{figure}

The normalized data $x_i(t)$ can be decomposed as (Fig.~\ref{fig:univariate_scheme})

\begin{equation}
x_i(t) = \sum_{k=1}^R \lambda_k a_k(t) u_{ik}, \label{eq:data_decomposition}
\end{equation}
where $a_k(t)$ is the $t$--th component of the $k$--th principal component $\mathbf{a}_k$ (PC) (temporal pattern) associated with the $k$--th EOF $\mathbf{u}_k$ (spatial pattern) with

\begin{equation}
a_k(t) = \sum_{j=1}^N u_{kj} x_j(t). \label{eq:pc}
\end{equation}
For many climatological data sets such as the precipitation and evaporation fields studied here, most of the variance in the data $\mathbf{x}(t)$ can be explained by a small number of EOFs, i.e., the eigenvalues $\lambda_k$ decay quickly with increasing rank $k$~(Fig.~\ref{fig:eigen_spectrum}). Equation~(\ref{eq:data_decomposition}) shows that in this situation, only a few EOFs and PCs are needed to closely approximate the data which allows the dimensionality reduction of high-dimensional data sets.

\subsection{Coupled pattern (maximum covariance) analysis}
\label{sec:coupled_patterns}

\begin{figure}[tb]
\begin{center}
\includegraphics[width=\columnwidth]{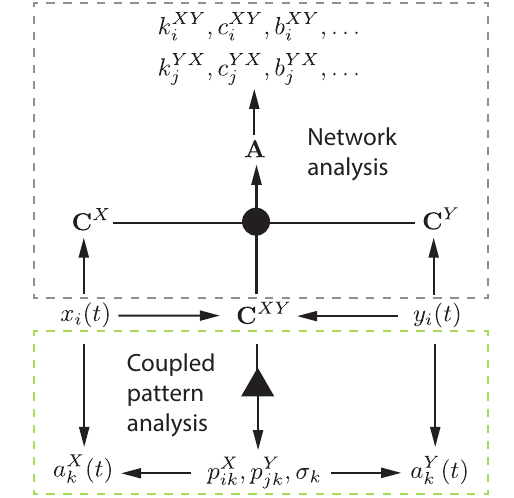}
\caption{A schematic outline of the relationship between coupled pattern (maximum covariance) and coupled climate network analysis in the spirit of the diagrams in \citet{Bretherton1992}. The singular value decomposition (SVD) operation is represented by the triangle, the thresholding operation by the disc. All vectors are written in component form.}
\label{fig:multivariate_scheme}
\end{center}
\end{figure}

Given two sets of normalized time series $\mathbf{x}(t) = \{x_i(t)\}_{i=1}^{N_X}$, and $\mathbf{y}(t) = \{y_j(t)\}_{j=1}^{N_Y}$ the \emph{cross-correlation matrix} $\mathbf{C}^{XY}=\{C_{ij}^{XY}\}_{ij}$ is defined by

\begin{equation}
C_{ij}^{XY} = \frac{1}{M} \sum_{t=1}^M x_i(t) y_j(t), \label{eq:cross_correlation_matrix}
\end{equation}
where $M$ is the length (number of samples) of each time series. $R$ in the following denotes the rank of $\mathbf{C}^{XY}$.

\begin{figure}[tb]
\begin{center}
\includegraphics[width=\columnwidth]{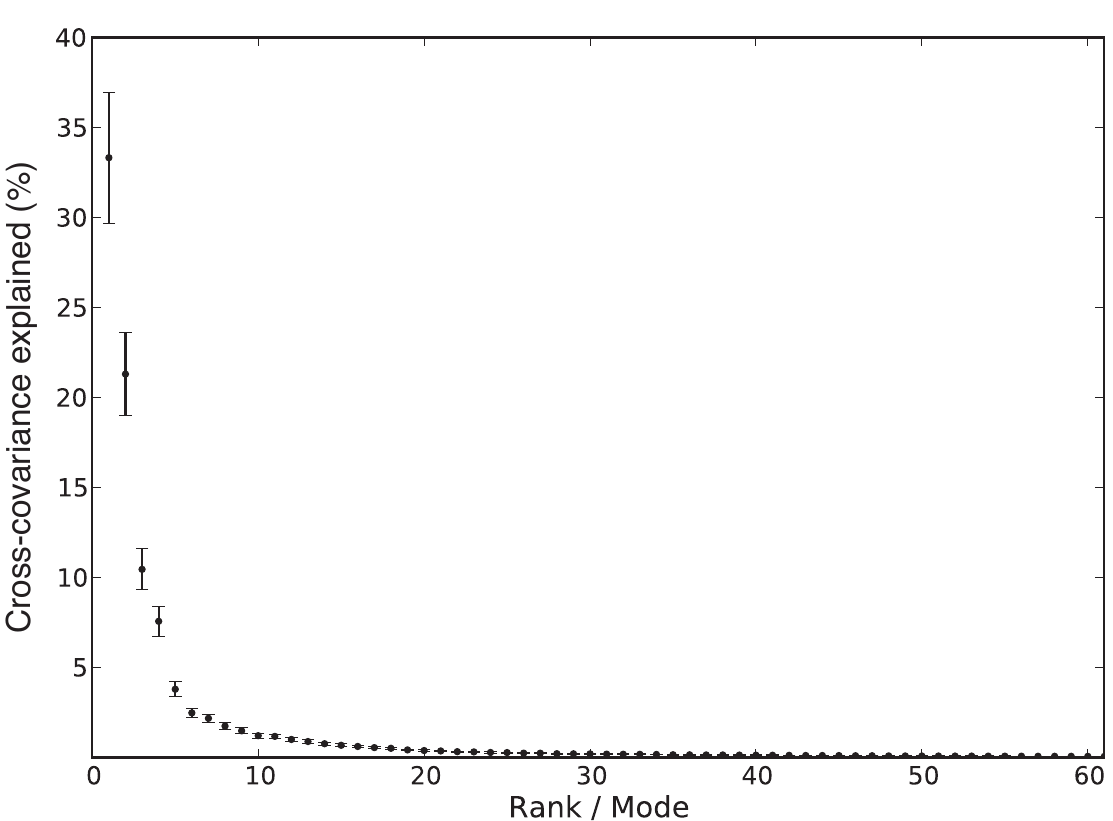}
\caption{Percentage of squared covariance $\sigma_k^2 / \sum_{l=1}^R \sigma_l^2$ between HOAPS-3 / GPCC precipitation ($X$) and HOAPS-3 evaporation ($Y$) data sets over the North Atlantic region (see Fig.~\ref{fig:svd_cross_deg_comp}) that is explained by pairs of coupled patterns $\mathbf{p}_k^X$, $\mathbf{p}_k^Y$. Most of the data sets' cross-covariance is captured by a small number of modes with the largest singular values $\sigma_k$. Error bars were estimated using North's rule of thumb~\citep{North1982}.}
\label{fig:svd_spectrum}
\end{center}
\end{figure}

Maximum covariance analysis identifies spatially ortho\-normal pairs of coupled patterns $\mathbf{p}_k^X = \{p_{ik}^X\}_{i=1}^{N_X}$, $\mathbf{p}_k^Y = \{p_{jk}^Y\}_{j=1}^{N_Y}$ that explain as much as possible of the temporal covariance between the two fields $\mathbf{x}(t)$ and $\mathbf{y}(t)$~\citep{Bretherton1992,vonStorch2003}. The coupled patterns can be found by solving the system of equations

\begin{align}
(\mathbf{C}^{XY})^T \mathbf{p}_k^X &= \sigma_k \mathbf{p}_k^Y \nonumber \\
\mathbf{C}^{XY} \mathbf{p}_k^Y &= \sigma_k \mathbf{p}_k^X
\end{align}
by means of a singular value decomposition of $\mathbf{C}^{XY}$ (Fig.~\ref{fig:multivariate_scheme}). Here, the $\mathbf{p}_k^X$ are an orthonormal set of $R$ vectors called \emph{left singular vectors}, the $\mathbf{p}_k^Y$ are an orthonormal set of $R$ vectors called \emph{right singular vectors}, and the $\sigma_k$ are non-negative numbers called \emph{singular values}, ordered such that $\sigma_1 \geq \sigma_2 \geq \dots \geq \sigma_R$. Here, $R$ denotes the rank of $\mathbf{C}^{XY}$. The total squared covariance explained by a certain pair of patterns $\mathbf{p}_k^X$, $\mathbf{p}_k^Y$ is $\sigma_k^2$. Therefore, the \emph{leading coupled patterns} $\mathbf{p}_1^X$, $\mathbf{p}_1^Y$ explain the largest fraction of squared covariance between the two fields of interest. In our example, taking into account only a few pairs of coupled patterns with the largest $\sigma_k$ already explains most of the covariance between the precipitation and evaporation fields~(Fig.~\ref{fig:svd_spectrum}).

The fields $\mathbf{x}(t),\mathbf{y}(t)$ can be expanded in terms of the coupled patterns as

\begin{align}
x_i(t) &= \sum_{k=1}^R a_k^X(t) p_{ik}^X, \\
y_i(t) &= \sum_{k=1}^R a_k^Y(t) p_{ik}^Y.
\end{align}
The expansion coefficients are obtained by projecting

\begin{align}
a_k^X(t) = \sum_{i=1}^R p_{ik}^X x_i(t), \\
a_k^Y(t) = \sum_{i=1}^R p_{ik}^Y y_i(t).
\end{align}

\section{Network techniques} \label{sec:network}

Complex network analysis offers a general framework for studying the structure of associations (links) between objects (nodes) that are of interest in many disciplines. Typical examples include the internet or world wide web in computer science, road networks and power grids in engineering, food webs in biology or social networks in sociology~\citep{Newman2003,Boccaletti2006,Newmanbook2010,Cohenbook2010}. It has become popular recently in several fields of science to apply the wealth of concepts and measures from complex network theory for the analysis of data that is even not given explicitly in network form. In network-based data analysis, a data set at hand, \textit{e.g.}, consisting of time series such as electroencephalogram, climate records, or spatiotemporal point events such as earthquake aftershock swarms, first has to be transformed to a network representation by means of a suitable algorithm or mathematical mapping. The resulting networks are referred to as \emph{functional networks} to distinguish them from \emph{structural networks} that are derived from systems with a more obvious graph structure, \textit{e.g.}, social networks or power grids. Examples of functional networks include gene regulatory networks in biology~\citep{Hempel2011}, functional brain networks in neuroscience~\citep{Bullmore2009}, CNs in climatology~\citep{Donges2009b,Donges2009a,Donges2011a}, or networks of earthquake aftershocks in seismology~\citep{Davidsen2008}. Forming a distinct class of methods, techniques for the network-based analysis of single or multiple time series such as recurrence networks \citep{Xu2008,Marwan2009,Donner2010b} and visibility graphs \citep{Lacasa2008} have recently been studied intensively with a focus on (paleo-)climatological applications~\citep{Donges2011b,Donges2011c,Hirata2011,Donner2012,Feldhoff2012a}.

The first functional network analysis of fields of climatological time series $\mathbf{x}(t)$ was presented by~\citet{Tsonis2004}, introducing the term \emph{climate network}\footnote{Note that the term \emph{climate network} is also used in distinct contexts that are unrelated to graph theory or data analysis, \textit{e.g.}, for describing collections of climatological/weather observation stations like the \emph{Greenland climate network}~\citep{Steffen2001} or associations of political organizations dealing with anthropogenic climate change such as the \emph{Climate Network Europe}~\citep{Raustiala2001}.}. Climate network analysis offers novel insights by transferring the toolbox of measures and algorithms from complex network theory to the study of climate system dynamics. Climate networks are simple graphs (i.e., there are no self-loops and at most one link between each pair of nodes) consisting of $N$ spatially embedded nodes $i$ that correspond to time series $x_i(t)$ representing observations, reanalyses, or simulations of climatological variables at fixed measurement stations, grid cells, or certain predefined regions. Links $\{i,j\}$ represent particularly strong or significant statistical interdependencies between two climate time series $x_i(t)$, $x_j(t)$, where usually a filtering procedure is applied first to reduce the effects of the annual cycle~\citep{Donner2008}.

Put differently, for a pairwise measure of statistical association $S_{ij}$ such as Pearson correlation~\citep{Tsonis2004,Tsonis2006}, mutual information~\citep{Donges2009a,Donges2009b,Palus2011}, transfer entropy~\citep{Runge2012}, or event synchronization~\citep{Malik2012,boers2013complex,Stolbova2014,boers2014complex}, a CN's adjacency matrix is given by

\begin{equation}
A_{ij} = \begin{cases}
    \Theta\left(S_{ij} - T_{ij}\right) & \text{if}\,\,\, i \neq j,\\
    0 & \text{otherwise},
\end{cases}
\end{equation}
where $\Theta(\cdot)$ is the Heaviside function, $T_{ij}$ denotes a threshold parameter, and $A_{ii}=0$ is set for all nodes $i$ to exclude self-loops. Usually, the threshold is fixed globally, \textit{i.e.}, $T_{ij}=T$ for all node pairs $(i,j)$. However, $T_{ij}$ may also be set for each pair individually to only include links with values of $S_{ij}$ exceeding a prescribed significance level, e.g., determined from a statistical test using surrogate time series~\citep{Palus2011}. In most studies, symmetric measures of statistical interdependency $S_{ij}=S_{ji}$ have been considered, leading to undirected CNs. However,~\citet{Gozolchiani2011}, \citet{Malik2012} and \citet{boers2014complex} exploited asymmetries in the cross-correlation function as well as in a measure of event synchronization to reconstruct directed CNs.

In the following, univariate and coupled CNs are introduced for studying the statistical interdependency structure within single fields as well as between two fields, respectively, together with graph-theoretical measures that are typically used for their quantification. For consistency with eigenanalysis (see Section~\ref{sec:eigen}), we restrict ourselves to linear Pearson correlation at zero lag as the measure of statistical association, \textit{i.e.}, $S_{ij}=|C_{ij}|$.

\subsection{Univariate climate networks}

Given a climatological field $\mathbf{x}(t)$, the \emph{adjacency matrix} $\mathbf{A}=\{A_{ij}\}_{ij}$ of the associated \emph{climate network} is given by

\begin{equation}
A_{ij} = \Theta(\left|C_{ij}^X\right| - T) - \delta_{ij} \label{eq:adjacency_matrix}
\end{equation}
with a prescribed global threshold $0 \leq T \leq 1$, where $\delta_{ij}$ denotes Kronecker's delta (see Eq.~(\ref{eq:correlation_matrix}) for the definition of $C_{ij}^X$). The absolute value of Pearson correlation $\left|C_{ij}^X\right|$ is commonly used, typically because negative correlations are considered equally important as positive ones~\citep{Tsonis2004}. Among others, univariate CNs have been studied by \citet{Tsonis2006,Tsonis2008a,Tsonis2008b,Yamasaki2008,Gozolchiani2008,Yamasaki2009,Donges2009b,Donges2009a,Tsonis2010,Berezin2011,Gozolchiani2011,Guez2011,Palus2011,Donges2011a,Tominski2011,Zou2011b,Malik2012,Rheinwalt2012,Rehfeld2013}.

\begin{figure}[htbp]
\begin{center}
\includegraphics[width=\columnwidth]{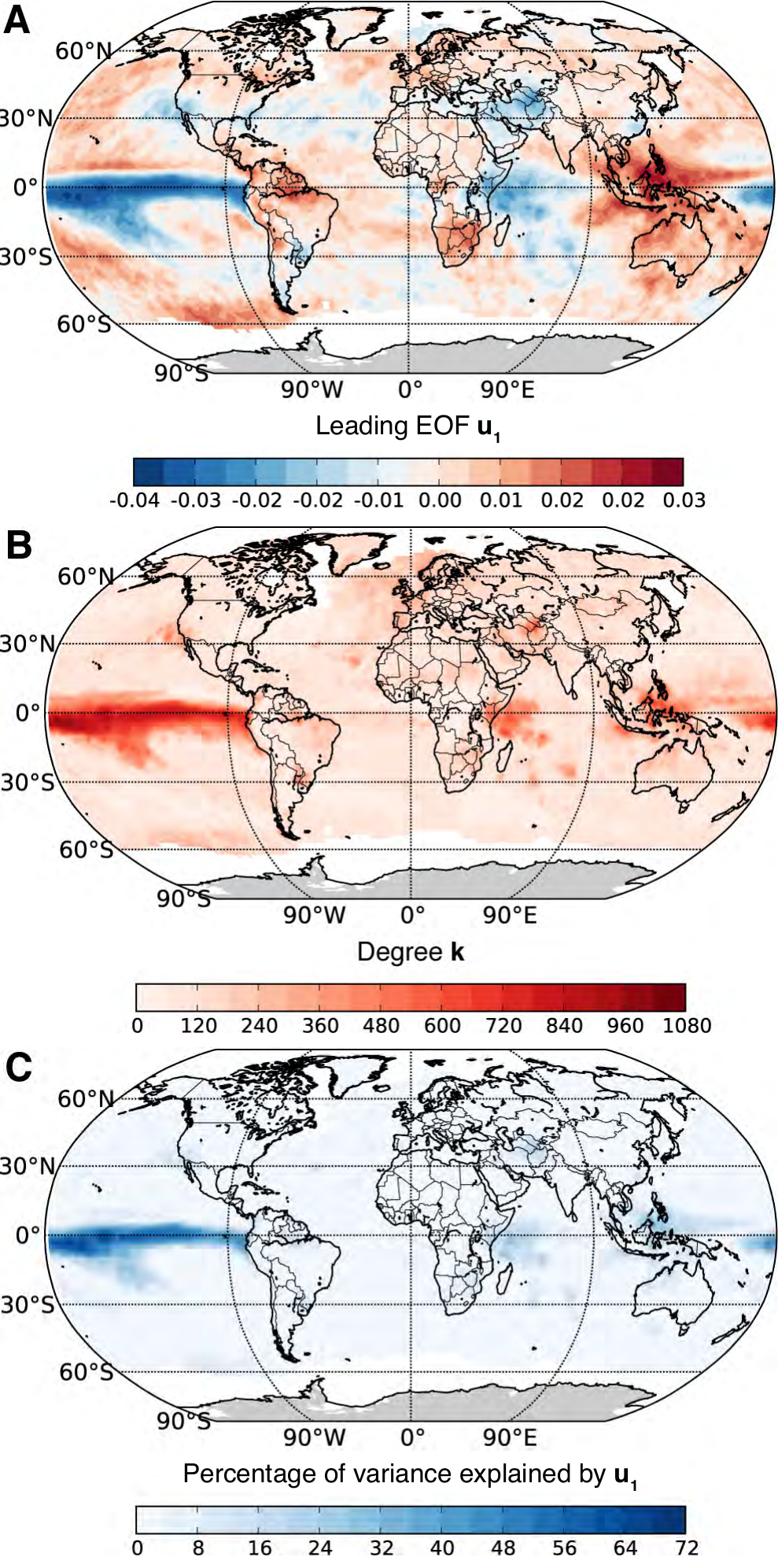}
\caption{Maps of (A)~first EOF $\mathbf{u}_1$, (B)~climate network degree field $\mathbf{k}$, and (C)~local percentage of variance explained by first EOF $\mathbf{u}_1$, $100 \times \textrm{Corr}(x_i(t),a_1(t))^2$ (homogeneous correlation map, see \citet{Bjornsson1997}), for the global HOAPS-3 / GPCC precipitation data set. The climate network construction threshold $T=0.27$ was chosen to yield a link density of $\rho=0.01$ (Eq.~(\ref{eq:link_density})). Note the similarity in the patterns displayed in panels (A)--(C) that is explained in Section~\ref{sec:relationships}.}
\label{fig:eof_degree_comp_precip}
\end{center}
\end{figure}

The \emph{degree} $k_i$ is the most frequently applied measure for studying CNs. It gives the number of network neighbors for each node $i$ and is defined as

\begin{align}
k_i &= \sum_{j=1}^N A_{ij} = \sum_{j=1}^N \Theta(\left|C_{ij}^X\right| - T) - 1. \label{eq:degree}
\end{align}
Maxima in the spatial pattern $\mathbf{k}$ with values of the degree that are much larger than average are referred to as \emph{super\-nodes} or \emph{hubs}~\citep{Tsonis2004,Tsonis2006}. These super-nodes indicate regions in the underlying field that are particularly strongly correlated to many other parts of the globe which are typically related to teleconnection patterns~\citep{Tsonis2008b}. For example, in the HOAPS-3 / GPCC precipitation data the most strongly connected region in the tropical Pacific (Fig.~\ref{fig:eof_degree_comp_precip}B) corresponds to the El Ni\~no-Southern Oscillation that is known to display global teleconnections~\citep{Ropelewski1987,Halpert1992,Tsonis2008b}.

Path-based centrality measures from network theory reveal higher-order patterns in the statistical interdependency structure of a climatological field~\citep{Donges2009b,Donges2009a,Palus2011}. High-order, in this context, refers to structures such as paths or network motifs that consist of two or more links, in contrast to the degree that is restricted to counting pairwise relationships between nodes. In this study, shortest-path closeness and betweenness are considered. \emph{Closeness centrality} $\mathbf{c} = \{c_i\}_{i=1}^N$ (CC) measures the inverse mean network distance of node $i$ to all other nodes via shortest paths and is defined as

\begin{equation}
c_i = \frac{N-1}{\sum_{j=1}^N l_{ij}},
\end{equation}
where $l_{ij}$ denotes the length of a shortest (or geodesic) path connecting nodes $i$ and $j$, \textit{i.e.}, the smallest number of links that are passed when traveling from $i$ to $j$ in the CN. In contrast, \emph{betweenness} $\mathbf{b} = \{b_i\}_{i=1}^N$ (BC) counts the relative number of shortest paths connecting any pair of nodes $j,k$ that include node $i$ and is defined as

\begin{equation}
b_i = \sum_{j=1}^N \sum_{k=1}^N \frac{n_{jk}(i)}{n_{jk}}.
\end{equation}
Here, $n_{jk}$ denotes the total number of shortest paths between $j,k$. $n_{jk}(i)$ gives the size of the subset of these paths that include $i$. CC and BC have been applied for comparing different types of CNs~\citep{Donges2009a}, revealing a backbone of energy flow in the surface air temperature field~\citep{Donges2009b}, unraveling the complex dynamics of the precipitation field during the Indian summer monsoon~\citep{Malik2012}, and studying the signatures of El Ni\~no and La Ni\~na events~\citep{Palus2011}. See Section~\ref{sec:benefits} for a more in depth discussion of the interpretation of these CN measures.

\subsection{Coupled climate networks}

\begin{figure}[hptb]
\centering
\resizebox{1.0\columnwidth}{!}{%
  \includegraphics{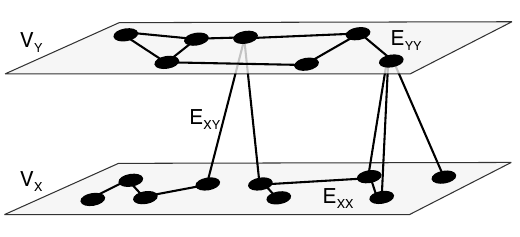}
}
\caption{A coupled climate network as it is constructed in this work, where $V_X$ and $V_Y$ denote the set of nodes in the subnetworks corresponding to grid points in data sets $\mathbf{x}(t)$ and $\mathbf{y}(t)$, respectively. $E_{XX}$ and $E_{YY}$ are sets of internal links within the subnetworks describing statistical relationships within each climatological field, while $E_{XY}$ contains information on their mutual statistical interdependencies. Figure is adapted from~\citep{Donges2011a}.}
\label{fig:coupled_network_sketch}
\end{figure}

One option for condensing information from more than one climatological observable in a CN is to define links based on statistical interdependencies between multivariate time series describing the dynamics of multiple observables rec\-orded at the same locations/nodes. For example,~\citet{Steinhaeuser2010} analyzed a CN constructed from surface air temperature, pressure, relative humidity, and precipitable water to extract regions of related climate variability. In contrast to this multivariate approach, coupled CNs are designed to represent statistical dependencies within and between two climatological fields $\mathbf{x}(t) = \{x_i(t)\}_{i=1}^{N_X}$, $\mathbf{y}(t) = \{y_j(t)\}_{j=1}^{N_Y}$ or within and between different regions~\citep{Donges2011a}. For this purpose, all time series from each of the involved climatological fields are associated to $N_X + N_Y$ nodes in the resulting network (Fig.~\ref{fig:coupled_network_sketch}). A coupled CN is defined by its adjacency matrix $\mathbf{A}$ that is obtained by thresholding the correlation matrix $\mathbf{C}$ of the concatenated fields $\mathbf{x}(t), \mathbf{y}(t)$, analogously to Eq.~(\ref{eq:adjacency_matrix}). Decomposing $\mathbf{C}$ as

\begin{equation}
\mathbf{C} = \begin{pmatrix} \mathbf{C}^X & \mathbf{C}^{XY} \\ (\mathbf{C}^{XY})^T & \mathbf{C}^Y \end{pmatrix}
\end{equation}
suggests to view coupled CNs as networks of networks or multilayer networks~\citep{Zhou2006,Buldyrev2010,Gao2011interdep_nets,Boccaletti2014}, where subnetworks (network layers) $G_X=(V_X,E_{XX})$ and $G_Y=(V_Y,E_{YY})$ are the induced subgraphs of the sets of nodes $V_X$, $V_Y$ belonging to data sets $\mathbf{x}(t)$, $\mathbf{y}(t)$, respectively (Fig.~\ref{fig:coupled_network_sketch}). While the edge sets $E_{XX}, E_{YY}$ describe the fields' internal correlation structure based on the correlation matrices $\mathbf{C}^X,\mathbf{C}^Y$, the set of cross-edges $E_{XY}$ captures dependencies between both fields and is based on the cross-correlation matrix $\mathbf{C}^{XY}$ (Fig.~\ref{fig:multivariate_scheme}). Coupled CNs have been applied for studying the Earth's atmosphere's general circulation structure~\citep{Donges2011a}, processes linking climate variability in the North Atlantic and North Pacific regions via the Arctic~\citep{Wiedermann2012,Wiedermann2014}, global atmosphere-ocean interactions~\citep{Feng2012}. Also, the coupled CN approach underlies the method developed in \citet{Ludescher2013,Ludescher2014} for forecasting El Ni\~no events.

\begin{figure*}[tbp]
\begin{center}
\includegraphics[width=0.75\textwidth]{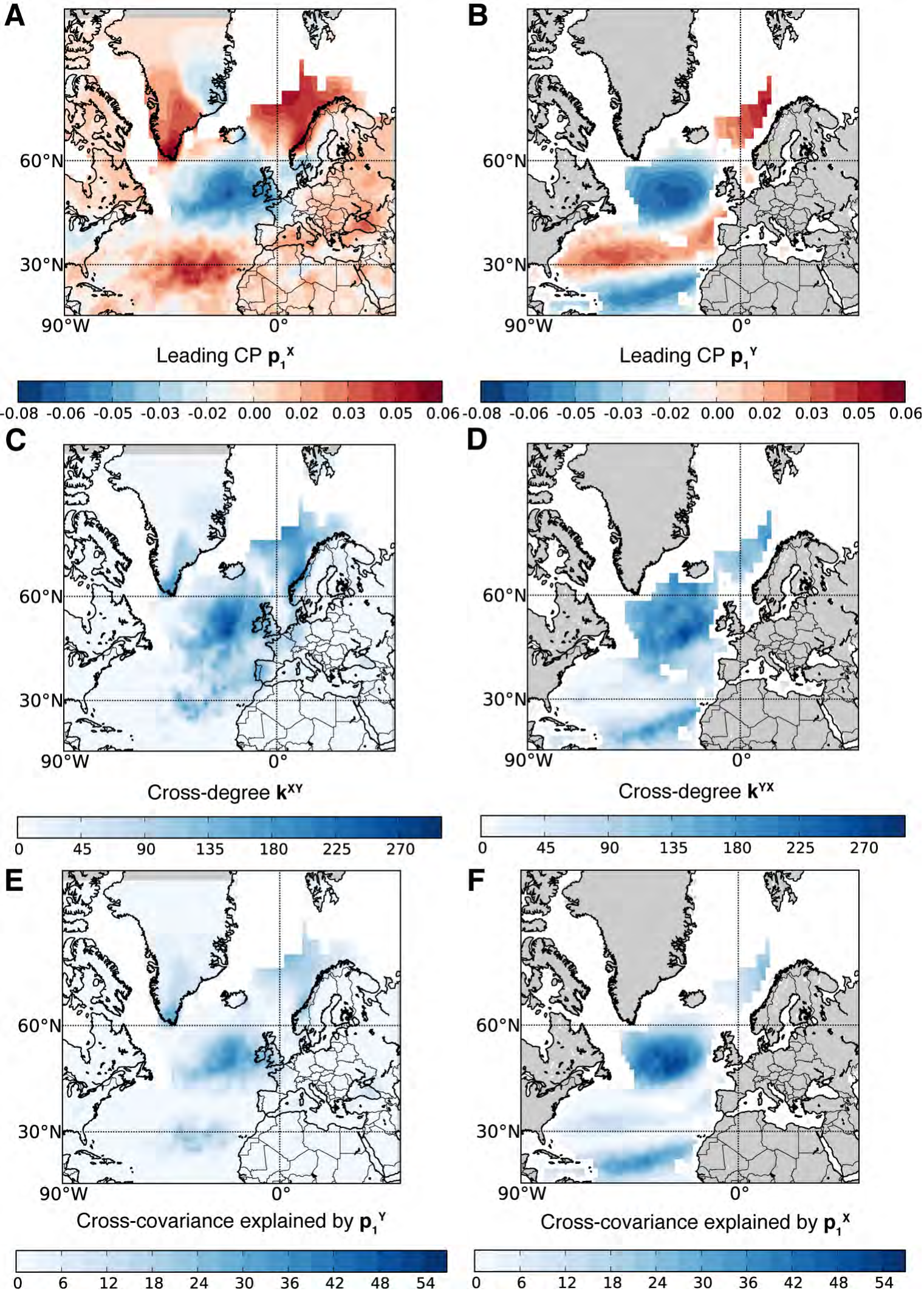}
\caption{Maps of leading pair of coupled patterns (A)~$\mathbf{p}_1^X$ and (B)~$\mathbf{p}_1^Y$, coupled climate network cross-degree fields (C)~$\mathbf{k}^{XY}$ and (D)~$\mathbf{k}^{YX}$, and percentage of cross-covariance explained by first pair of coupled patterns (E)~$\mathbf{p}_1^Y$, $100 \times \textrm{Corr}(x_i(t),a_1^Y(t))^2$, and (F)~$\mathbf{p}_1^X$, $100 \times \textrm{Corr}(y_i(t),a_1^X(t))^2$ (heterogeneous correlation maps, see \citet{Bjornsson1997}), for the HOAPS-3 / GPCC precipitation ($X$) and HOAPS-3 evaporation ($Y$) data sets over the North Atlantic. For constructing the coupled climate network, a threshold $T=0.47$ was chosen to yield a cross-link density of $\rho_{XY}=0.01$ (Eq.~(\ref{eq:cross_link_densities})) resulting in internal link densities $\rho_X=0.01$ and $\rho_Y=0.06$~\citep{Donges2011a}.}
\label{fig:svd_cross_deg_comp}
\end{center}
\end{figure*}

The statistical interdependency structure between fields $\mathbf{x}(t)$, $\mathbf{y}(t)$ can be quantified with a set of graph-theoretical measures developed for investigating the topology of networks of interacting networks \citep{Donges2011a}. The \emph{cross-degree} $\mathbf{k}^{XY} = \{k_i^{XY}\}_{i=1}^{N_X}$ is the number of neighbors of node $i \in V_X$ in subnetwork $G_Y$:

\begin{align}
k_i^{XY} &= \sum_{j \in V_Y} A_{ij} = \sum_{j=1}^{N_Y} A_{ij}^{XY} = \sum_{j=1}^{N_Y} \Theta(|C_{ij}^{XY}| - T). \label{eq:cross_degree}
\end{align}
Analogously, the cross-degree $\mathbf{k}^{YX} = \{k_j^{YX}\}_{j=1}^{N_Y}$ is given by

\begin{align}
k_j^{YX} &= \sum_{i \in V_X} A_{ij} = \sum_{i=1}^{N_X} A_{ij}^{XY} = \sum_{i=1}^{N_X} \Theta(|C_{ij}^{XY}| - T).
\end{align}
Similarly to degree in univariate climate net\-works, regions $i$ in field $\mathbf{x}(t)$ with a large cross-degree $k_i^{XY}$ are considered to be strongly dynamically interrelated with many locations in field $\mathbf{y}(t)$ and vice versa. For the precipitation and evaporation data sets (Fig.~\ref{fig:svd_cross_deg_comp}C,D), such regions with high cross-connectivity correspond to major covariability areas of evaporation and precipitation fields driven by the North-Atlantic Oscillation (NAO) \citep{Andersson2010,Petrova2012}.

Furthermore, analogously to univariate climate net\-works, generalizations of path-based measures for network of networks can be derived~\citep{Donges2011a}. Here, cross-closeness and cross-betweenness are considered.  \emph{Cross-close\-ness} $\mathbf{c}^{XY} = \{c_i^{XY}\}_{i=1}^{N_X}$ (cross-CC) measures the inverse mean network distance of node $i \in V_X$ to all nodes $j \in V_Y$ via shortest paths and is defined as

\begin{equation}
c_i^{XY} = \frac{N_X + N_Y -1}{\sum_{j \in V_Y} l_{ij}}.
\end{equation}
\emph{Cross-betweenness} $\mathbf{b}^{XY} = \{b_i^{XY}\}_{i=1}^{N_X}$ (cross-BC) counts the relative number of shortest paths connecting any pair of nodes $j \in V_X, k \in V_Y$ that include node $i \in V_X$ and is defined as

\begin{equation}
b_i^{XY} = \sum_{j \in V_X} \sum_{k \in V_Y} \frac{n_{jk}(i)}{n_{jk}}.
\end{equation}
For nodes $j$ in field $\mathbf{y}(t)$, the measures $\mathbf{c}^{YX} = \{c_j^{YX}\}_{j=1}^{N_Y}$ and $\mathbf{b}^{YX} = \{b_j^{YX}\}_{j=1}^{N_Y}$ are obtained from analogous expressions following~\citet{Donges2011a}. Interpretations of coupled CN measures will be discussed in Section~\ref{sec:benefits}.

\section{Relationships between eigen and climate network analysis} \label{sec:relationships}

Comparing the results of eigen and CN analysis, notable similarities become apparent, e.g., in the leading EOF $\mathbf{u}_1$ and CN degree $\mathbf{k}$ for the HOAPS-3 / GPCC precipitation data (Fig.~\ref{fig:eof_degree_comp_precip}). Analogous relations are observed when inspecting leading coupled patterns and coupled CN cross-degree for HOAPS-3 / GPCC precipitation and HOAPS-3 evaporation data (Fig.~\ref{fig:svd_cross_deg_comp}). To explain these similarities, in this section, formal relationships between patterns from eigen and CN analysis are derived and illustrated empirically for global precipitation and evaporation data sets.  Relations between single field (EOFs and univariate CN measures, Section~\ref{sec:relations_single_field}) as well as multiple field patterns (coupled patterns and coupled CN measures, Section~\ref{sec:relations_coupled_field}), and temporal patterns are discussed. Note that similar relationships hold when both eigen and network analysis are based on a type of symmetric similarity matrix that is different from linear correlation at zero lag, \textit{e.g.}, considering mutual information~\citep{Donges2009b,Donges2009a} or the ISOMAP algorithm~\citep{Tenenbaum2000,Gamez2004}.

\subsection{Single field patterns} \label{sec:relations_single_field}

As the correlation matrix $\mathbf{C}^X$ is symmetric and, hence, diagonalizable, it can be decomposed with respect to its eigensystem such that

\begin{equation}
C_{ij}^X = \sum_{k=1}^R u_{ik} \lambda_k u_{jk}. \label{eq:eigen_decomp}
\end{equation}
If the leading EOF $\mathbf{u}_1$ explains a large fraction of the total variance, \textit{i.e.}, if $\lambda_1 \gg \lambda_2$, then $C_{ij}^X$ can be approximated as

\begin{equation}
C_{ij}^X \approx \lambda_1 u_{i1} u_{j1}.
\end{equation}
Inserting this expression into the definition of CN degree (Eq.~(\ref{eq:degree})) yields

\begin{equation}
k_i \approx \sum_{j=1}^N \Theta(\lambda_1 \left|u_{i1} u_{j1}\right| - T) - 1. \label{eq:degree_approx}
\end{equation}
This approximation explains the empirically observed similarity between degree $\mathbf{k}$ and the leading EOF $\mathbf{u}_1$ (compare Fig.~\ref{fig:eof_degree_comp_precip}, panels A and B, for the precipitation data set) in the following way: All nodes $j$ with $|u_{j1}| > \frac{T}{\lambda_1 |u_{i1}|}$ contribute to the degree $k_i$ at node $i$, hence, a larger $|u_{i1}|$ typically leads to more positive contributions to the sum in Eq.~(\ref{eq:degree_approx}) and, therefore, to a larger degree $k_i$. Consequently, CN degree $\mathbf{k}$ and the vector of absolute values of the leading EOF's elements $|\mathbf{u}_1|$ are expected to be positively correlated.

\begin{figure}[tbp]
\begin{center}
\includegraphics[width=\columnwidth]{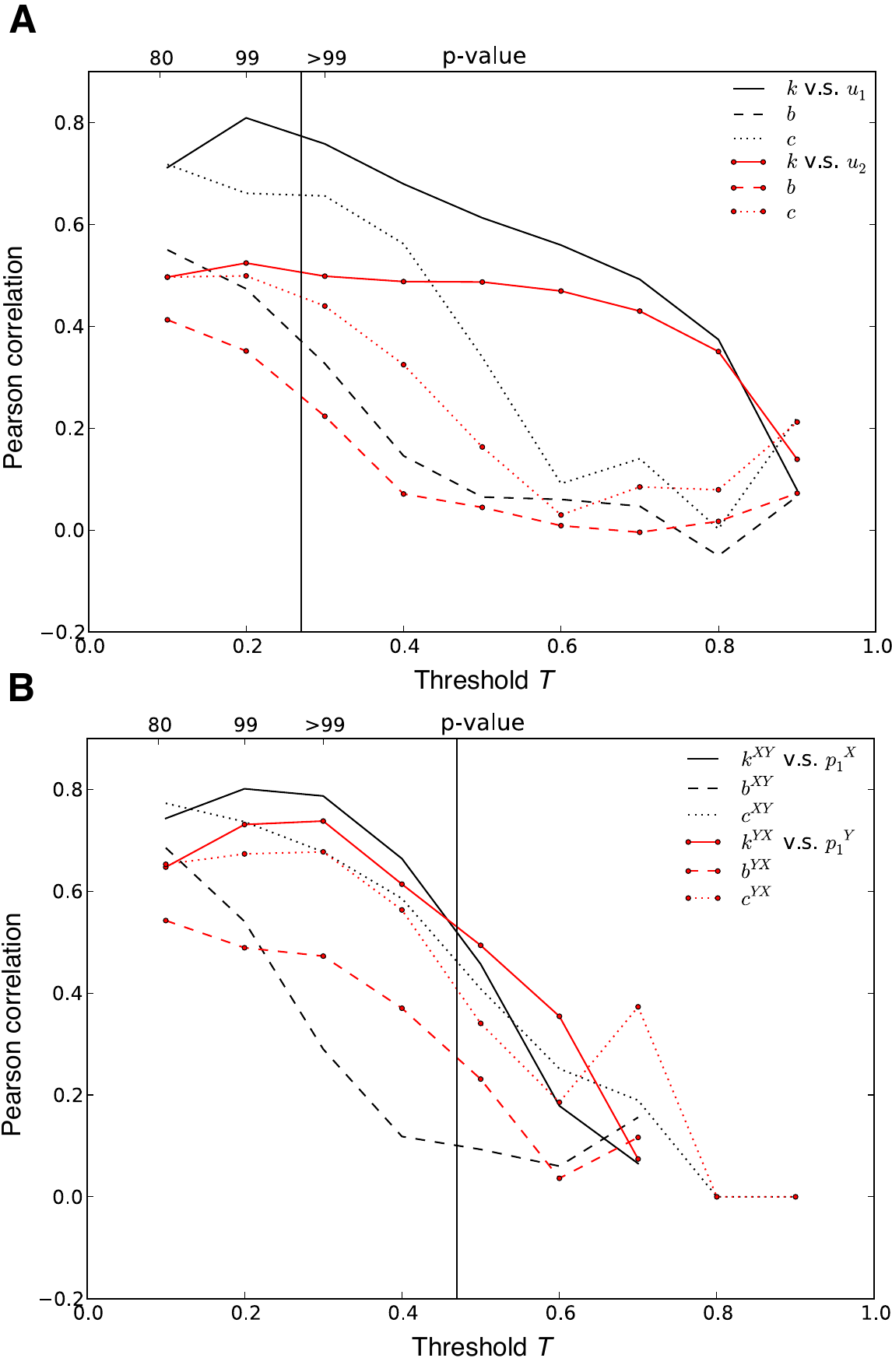}
\caption{Linear correlations between spatial patterns from eigen and network techniques for climate data analysis. Pearson correlation between (A)~the absolute values of the first two EOFs $|\mathbf{u}_1|,|\mathbf{u}_2|$ and CN measures degree $\mathbf{k}$, closeness $\mathbf{c}$ and betweenness $\mathbf{b}$ for HOAPS-3 / GPCC precipitation data as well as (B)~the first coupled patterns $\mathbf{p}_1^X, \mathbf{p}_1^Y$ and coupled CN measures cross-degree $\mathbf{k}^{XY},\mathbf{k}^{YX}$, cross-closeness $\mathbf{c}^{XY},\mathbf{c}^{YX}$, and cross-betweenness $\mathbf{b}^{XY},\mathbf{b}^{YX}$ for HOAPS-3 / GPCC precipitation (X) and HOAPS-3 evaporation data. In both panels, correlations are displayed for varying network construction threshold $T$, where the corresponding $p$-value according to the Student's $t$-test is given on the upper horizontal axis. Vertical lines in panels (A) and (B) indicate the thresholds used in Figs.~\ref{fig:eof_degree_comp_precip} and \ref{fig:svd_cross_deg_comp}, respectively.}
\label{fig:pattern_correlation}
\end{center}
\end{figure}

For the global precipitation data set, a large positive correlation between $\mathbf{k}$ and $|\mathbf{u}_1|$ is indeed detected for intermediate thresholds $T$ of the order where CNs are typically constructed~\citep{Donges2009a}, while for smaller and larger thresholds, the correlation decreases (Fig.~\ref{fig:pattern_correlation}A). The latter is expected, since both for $T \to 0$ (fully connected network) and $T \to 1$ (network devoid of links), the CN contains no information about the climatological field anymore and the degree field is constant with $k_i \to N-1$ and $k_i \to 0$ for all nodes $i$, respectively. Hence, maximum pattern correspondence is expected for intermediate thresholds $T$ (for these as well as computational reasons, results for $T=0$ and $T=1$ are not included in Fig.~\ref{fig:pattern_correlation}). Notably, selecting $T$ as maximizing the correlation between degree $\mathbf{k}$ and the leading EOF $|\mathbf{u}_1|$ could provide a criterion for an informed choice of the threshold $T$. Such a choice would approximate a situation where the information that the CN contains on linear statistical interdependencies in the field of interest is maximized. Further work is needed to develop more suitable criteria for defining binary CNs with maximum information content. Furthermore and as expected, the correlation between degree $\mathbf{k}$ and the second EOF $|\mathbf{u}_2|$ is mostly smaller than that between degree and leading EOF (Fig.~\ref{fig:pattern_correlation}A).

Using the full eigen-decomposition of $\mathbf{C}^X$, an exact relationship between the degree $\mathbf{k}$ and all EOFs $\mathbf{u}_k$ together with their associated eigenvalues $\lambda_k$ can be derived as

\begin{equation}
k_i =  \sum_{j=1}^N \Theta\left(\left|\sum_{k=1}^R u_{ik} \lambda_k u_{jk}\right| - T\right) - 1. \label{eq:degree_exact}
\end{equation}
Using this expression, the scalar link density

\begin{align}
\rho=\frac{\left<k_i\right>_{i=1}^N}{N-1} \label{eq:link_density}
\end{align}
can likewise be expanded or approximated, where $\left<\cdot\right>$ denotes the arithmetic mean.
Similarly, a relationship between area-weighted EOFs~\citep{Hannachi2007}, the area-weighted degree~\citep{Heitzig2012} (also called area weighted connectivity~\citep{Tsonis2006}) and all other network measures directly expressible in terms of the adjacency matrix $A_{ij}$ can be derived.

\subsection{Coupled patterns} \label{sec:relations_coupled_field}

The cross-correlation matrix $\mathbf{C}^{XY}$ can be decomposed in terms of singular values and coupled patterns as (Fig.~\ref{fig:multivariate_scheme})

\begin{equation}
C_{ij}^{XY} = \sum_{k=1}^R \sigma_k p_{ik}^X p_{jk}^Y.
\end{equation}
The relationship between cross-degree $\mathbf{k}^{XY}$, $\mathbf{k}^{YX}$ and coupled patterns $\mathbf{p}_k^X$, $\mathbf{p}_k^Y$  can then be derived as above:

\begin{align}
k_i^{XY} &= \sum_{j=1}^{N_Y} \Theta\left(\left|\sum_{k=1}^R \sigma_k p_{ik}^X p_{jk}^Y\right| - T\right) \label{eq:cross_degree_exact_1} \\
& \approx \sum_{j=1}^{N_Y} \Theta\left(\sigma_1 \left|p_{i1}^X p_{j1}^Y\right| -T\right),\\
k_j^{YX} &= \sum_{i=1}^{N_X} \Theta\left(\left|\sum_{k=1}^R \sigma_k p_{ik}^X p_{jk}^Y\right| -T\right) \label{eq:cross_degree_exact_2} \\
& \approx \sum_{i=1}^{N_X} \Theta\left(\sigma_1 \left|p_{i1}^X p_{j1}^Y\right| -T\right).
\end{align}
The approximations hold for the maximum singular value fulfilling $\sigma_1\gg\sigma_2 \geq \dots \geq \sigma_R$. $R$ is the rank of the cross-correlation matrix $\mathbf{C}^{XY}$. By a similar argument as given above this shows that $\mathbf{k}^{XY}$ and $|\mathbf{p}_1^X|$ ($\mathbf{k}^{YX}$ and $|\mathbf{p}_1^Y|$) are expected to be positively correlated which is consistent with our results regarding the interdependency structure between precipitation and evaporation fields. While in our example, the correspondence between the resulting patterns is somewhat less pronounced than in the single-field setting (Fig.~\ref{fig:pattern_correlation}B), still regions with a strongly negative loading in the leading coupled patterns $\mathbf{p}_1^X$ and $\mathbf{p}_1^Y$ appear as super nodal structures in the cross-degree fields (Fig.~\ref{fig:svd_cross_deg_comp}). When studying varying network construction thresholds $T$, as in the case of single-field patterns, the correlation between the absolute values of the leading pair of coupled patterns and cross-degree fields is maximum for intermediate $T$ and decreases for $T \to 0$ and $T \to 1$ (Fig.~\ref{fig:pattern_correlation}B). Also, consistently with Eqs.~(\ref{eq:cross_degree_exact_1}) and (\ref{eq:cross_degree_exact_2}), the correlation between the second pair of coupled patterns and cross-degree fields is always smaller than that observed for the leading pair of coupled patterns (results not shown).

The scalar cross-link densities~\citep{Donges2011a}

\begin{align}
\rho_{XY}=\frac{\left<k^{XY}_i\right>_{i=1}^{N_X}}{N_Y} \nonumber \\
\rho_{YX}=\frac{\left<k^{YX}_j\right>_{j=1}^{N_Y}}{N_X} \label{eq:cross_link_densities}
\end{align}
can also be expanded and approximated in terms of CPs and singular values using the above expressions. Analogously, area-weighted coupled patterns~\citep{vonStorch2003} are related to the area-weighted cross-degree introduced by \citet{Feng2012} and \citet{Wiedermann2012}.

\subsection{Temporal patterns}

In EOF analysis, temporal patterns (principal components) $a_k(t)$ describing the evolution of their associated spatial patterns $\mathbf{u}_k$ are easily obtained by projecting the data $\mathbf{x}(t)$ onto the latter patterns $\mathbf{u}_k$ (Eq.~(\ref{eq:pc})). Analogously, the same holds for multivariate extensions such as coupled pattern analysis~\citep{Bretherton1992,vonStorch2003}, see Section~\ref{sec:eigen}. In CN analysis, however, the temporal evolution of spatial network measure patterns such as the degree $\mathbf{k}$ or betweenness $\mathbf{b}$ cannot be directly obtained from the adjacency matrix $\mathbf{A}$ and $\mathbf{x}(t)$. To allow the study of non-stationarities in the statistical interdependence structure of climatological fields, several authors have investigated the evolving local (\textit{e.g.}, $\mathbf{k}(t)$ or $\mathbf{b}$(t)) and global properties of CNs $\mathbf{A}(t)$ constructed from temporal windows sliding over the time series data~\citep{Gozolchiani2008,Yamasaki2008,Yamasaki2009,Gozolchiani2011,Guez2011,Berezin2011,Carpi2012,Martin2013,Radebach2013,Ludescher2013,Ludescher2014}. A similar strategy could be applied to coupled CN analysis.

It should be noted that unlike in the above sections, no direct relationship can be derived linking temporal patterns from eigen and network analysis. The reason for this is two\-fold. First, temporal patterns $a_k(t)$ of standard EOF analysis are based on the full data set $\mathbf{x}(t)$, while the evolving spatial network patterns are computed from subsets (defined by temporal windows) of $\mathbf{x}(t)$. Second, since temporal patterns $a_k(t)$ of eigenanalysis are merely scalar prefactors in the expansion Eq.~(\ref{eq:data_decomposition}) (see Figs.~\ref{fig:univariate_scheme} and \ref{fig:multivariate_scheme}), the spatial EOF patterns $\mathbf{u}_k$ are time-independent, whereas evolving CN measures such as $\mathbf{k}(t)$ can vary independently at every location $i$. Hence, in contrast to standard EOF patterns, the spatial patterns in the network properties derived from evolving CNs are explicitly time-dependent. The latter case is analogous to extended EOF analysis, where standard EOF analysis is applied in a sliding-window mode as well \citep{Fraedrich1997}.

\section{Discussion} \label{sec:benefits}

The relationships derived in the previous section provide guidance on deciding how and in which applications CN analysis can be expected to yield information that is complementary to the results of eigenanalysis. Particularly, we will focus on a discussion and climatological interpretation of single field and coupled patterns derived from precipitation and evaporation data~(Section~\ref{sec:disc_precip_evap}) and relate this to a study of single field patterns for global surface air temperature data~(Section~\ref{sec:disc_sat}). Based on these insights, we point out some methodological as well as practical potentials of CN analysis of climatological fields (Section~\ref{sec:disc_potentials}).

\subsection{Precipitation and evaporation data} \label{sec:disc_precip_evap}

For the HOAPS-3 / GPCC precipitation and HOAPS-3 evaporation data sets, pronounced similarities between the features observed in the degree or cross-degree fields and those in the leading EOF or coupled patterns that are derived from the same data have been described and explained mathematically (Section~\ref{sec:relationships}). 
More specifically, active regions displaying strong correlations with many other locations, and, hence, a large degree or cross-degree (termed super-nodes in the context of CN analysis~\citep{Tsonis2004,Tsonis2006,Barreiro2011}) correspond to regions with large positive or negative loading in the leading EOF or coupled patterns. For example, this can be observed for the equatorial Pacific in the precipitation data (Fig.~\ref{fig:eof_degree_comp_precip}A,B). The spatial similarity between the amplitude of the leading EOF and CN degree field reveals the well-known ENSO variability pattern~\citep{Ropelewski1987}. Particularly, the patterns in the explained variance fraction (Fig.~\ref{fig:eof_degree_comp_precip}C) closely resemble high connectivity areas of the CN resembling most prominent ENSO teleconnections \citep{Andersson2010,Halpert1992,Ropelewski1987}. Additional dipole information described by the EOF is typically preserved by neighbors of the network's major super-nodes (not shown here, see \citet{Petrova2012} and \citet{Kawale2013}).

Considering the bivariate analysis of precipitation and evaporation data over North Atlantic (Fig.~\ref{fig:svd_cross_deg_comp}), regions with a strongly negative loading in the leading pair of coupled patterns appear as super nodal structures in the cross-degree fields obtained from coupled CN analysis. Areas with a high fraction of explained cross-covariance (Fig.~\ref{fig:svd_cross_deg_comp}E,F) well correspond to the coupled network topology as indicated by the cross-degree fields (Fig.~\ref{fig:svd_cross_deg_comp}C,D) and all together depict major covariability areas of evaporation and precipitation driven by the NAO. The cross-degree field $\mathbf{k}^{XY}$ (Fig.~\ref{fig:svd_cross_deg_comp}C), displaying the number of strong correlations between precipitation variability at a certain location with evaporation dynamics at all other grid points, reveals teleconnections associated to the NAO over the southern tip of Greenland as well as a positive NAO signal over Portugal and a negative NAO signal over Norway~\citep{Andersson2010}. In turn, the cross-degree field $\mathbf{k}^{YX}$ (Fig.~\ref{fig:svd_cross_deg_comp}D), showing the number of strong correlations between evaporation dynamics at one point and precipitation variability at all other locations, is only available over the ocean and follows the covariance structure of the main evaporation
determinant parameters with NAO~\citep{Cayan1992,Marshall2001}.


\begin{figure}[hptb]
\begin{center}
\includegraphics[width=\columnwidth]{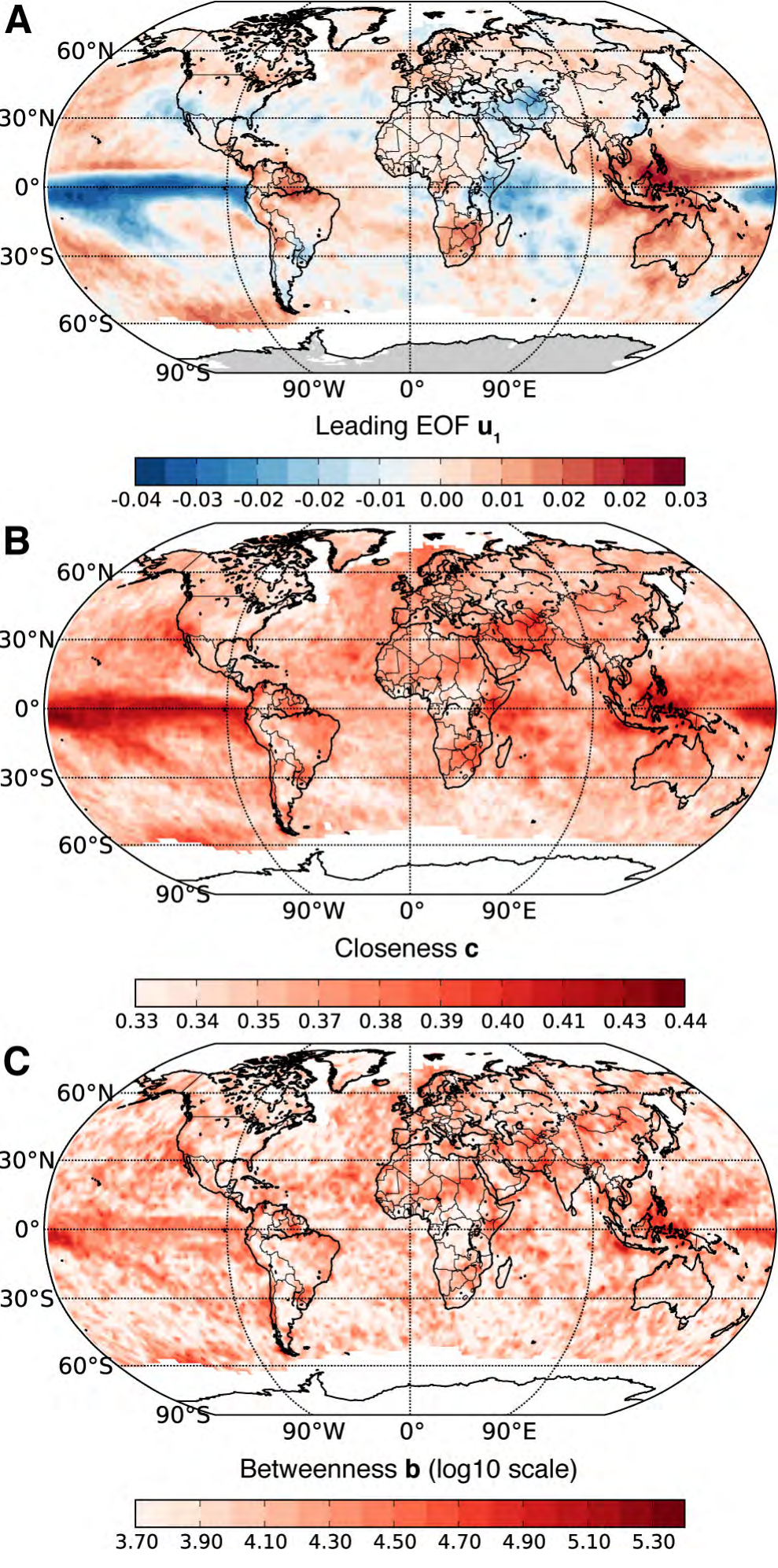}
\caption{Maps of (A) leading EOF $\mathbf{u}_1$, (B)~closeness field $\mathbf{c}$, and (C)~betweenness field $\mathbf{b}$ for the global HOAPS-3 / GPCC precipitation climate network. The network construction threshold $T=0.27$ was chosen to yield a link density of $\rho=0.01$.}
\label{fig:path_based_measures}
\end{center}
\end{figure}

Beyond the frequently studied degree $\mathbf{k}$, complex network theory provides a wealth of additional measures that can be used to study higher-order properties of the statistical interdependency structure within and between climatological fields. For example, the mentioned measures based on the properties of shortest paths in (coupled) CNs such as (cross-) closeness $\mathbf{c}$ ($\mathbf{c}^{XY}, \mathbf{c}^{YX}$) and (cross-) betweenness $\mathbf{b}$ ($\mathbf{b}^{XY}, \mathbf{b}^{YX}$) (Fig.~\ref{fig:path_based_measures}) have been argued to give insights on the local speed of propagation as well as the preferred pathways for the spread of perturbations within or between the studied fields, respectively~\citep{Donges2009b,Donges2009a,Donges2011a,Malik2012,Molkenthin2014a}. In this way, CN analysis has the potential to unveil information on climate dynamics from climatological field data that conceptually supplements the results of eigenanalysis.

Focussing on the precipitation data to further investigate this aspect, we find that the correlation of CC and BC to the first two EOFs obtained from the data are systematically and significantly smaller than that between the degree field and the same EOFs (Fig.~\ref{fig:pattern_correlation}A). Similarly, in the bivariate case, the correlations of cross-CC and cross-BC with the leading coupled pattern are considerably smaller than those between the latter and cross-degree for most thresholds $T$ (Fig.~\ref{fig:pattern_correlation}B). However, for the HOAPS-3 / GPCC precipitation data, the patterns observed in the leading EOF resemble those found in the CC and BC fields (Fig.~\ref{fig:path_based_measures}) as well as those in the degree field (Fig.~\ref{fig:eof_degree_comp_precip}). These results can be explained from a network point of view by considering that precipitation fields are typically only correlated on short spatial scales and display a smaller degree of spatial coherency when compared to other atmospheric variables such as pressure or temperature~\citep{Feldhoff2014}. In turn, this leads to a larger degree of randomness in the structure of CNs constructed from this data. In random networks, correlations between centrality measures such as degree, closeness and betweenness arise \citep{Boccaletti2006}. In other words, spatially incoherent climatological fields can give rise to CNs with a notable degree of disorder in the placement of links between different nodes which induces correlations between network centrality measures. For the precipitation data set at hand, the first eigenvalue separates from the remaining spectrum (Fig.~\ref{fig:eigen_spectrum}) leading to a pronounced correlation between the leading EOF $\mathbf{u}_1$ and the degree field (see Eq.~\ref{eq:degree_approx}), and, hence, to correlations between $\mathbf{u}_1$ and CC, BC.

\subsection{Surface air temperature data} \label{sec:disc_sat}

\begin{figure}[tb]
\begin{center}
\includegraphics[width=\columnwidth]{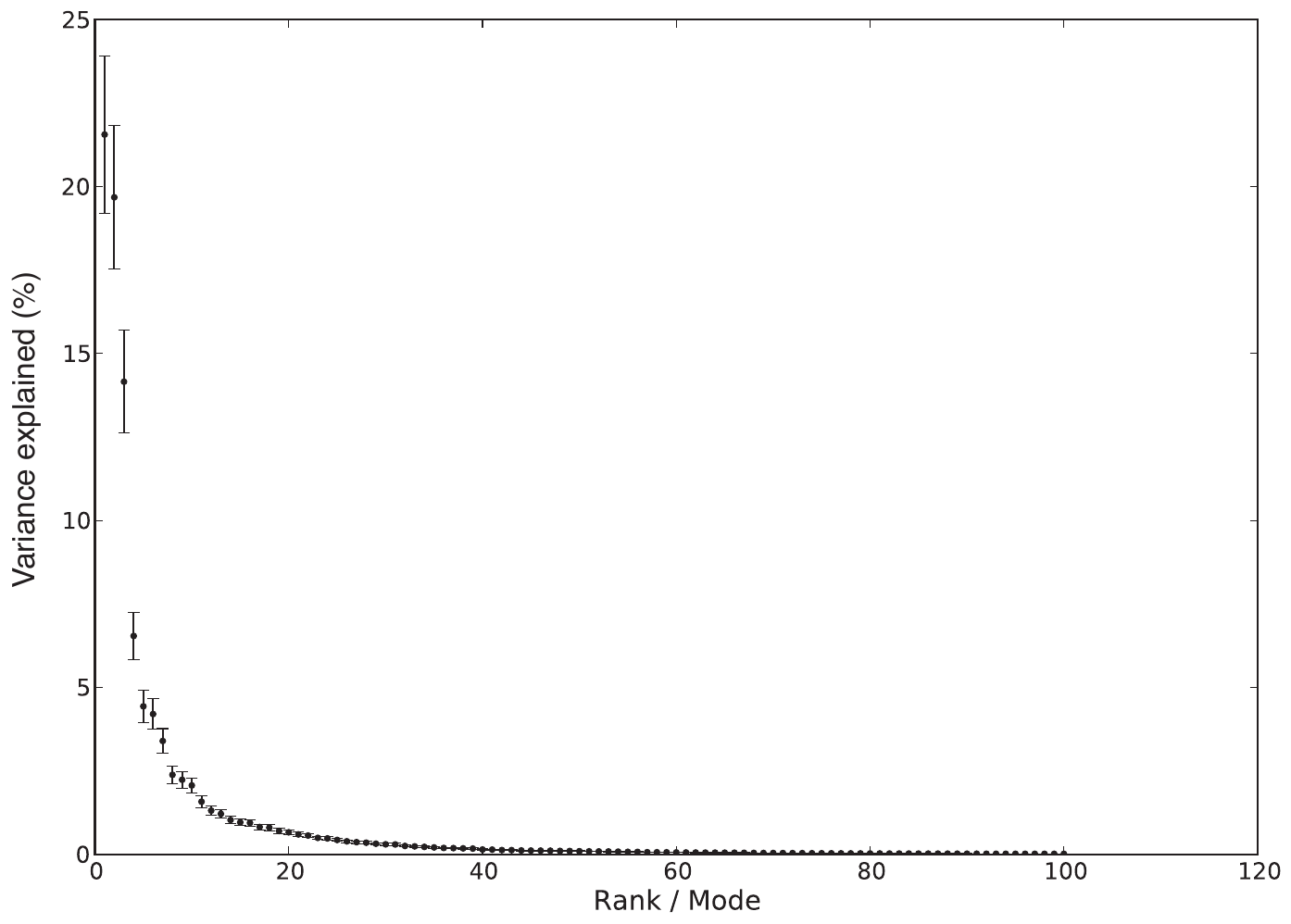}
\caption{Percentage of variance $\lambda_k / \sum_{l=1}^R \lambda_l$ explained by EOFs $\mathbf{u}_k$ for the NCEP/NCAR surface air temperature data set. Error bars were estimated using North's rule of thumb~\citep{North1982}.}
\label{fig:eigen_spectrum_sat}
\end{center}
\end{figure}

\begin{figure*}[tbp]
\begin{center}
\includegraphics[width=0.95\textwidth]{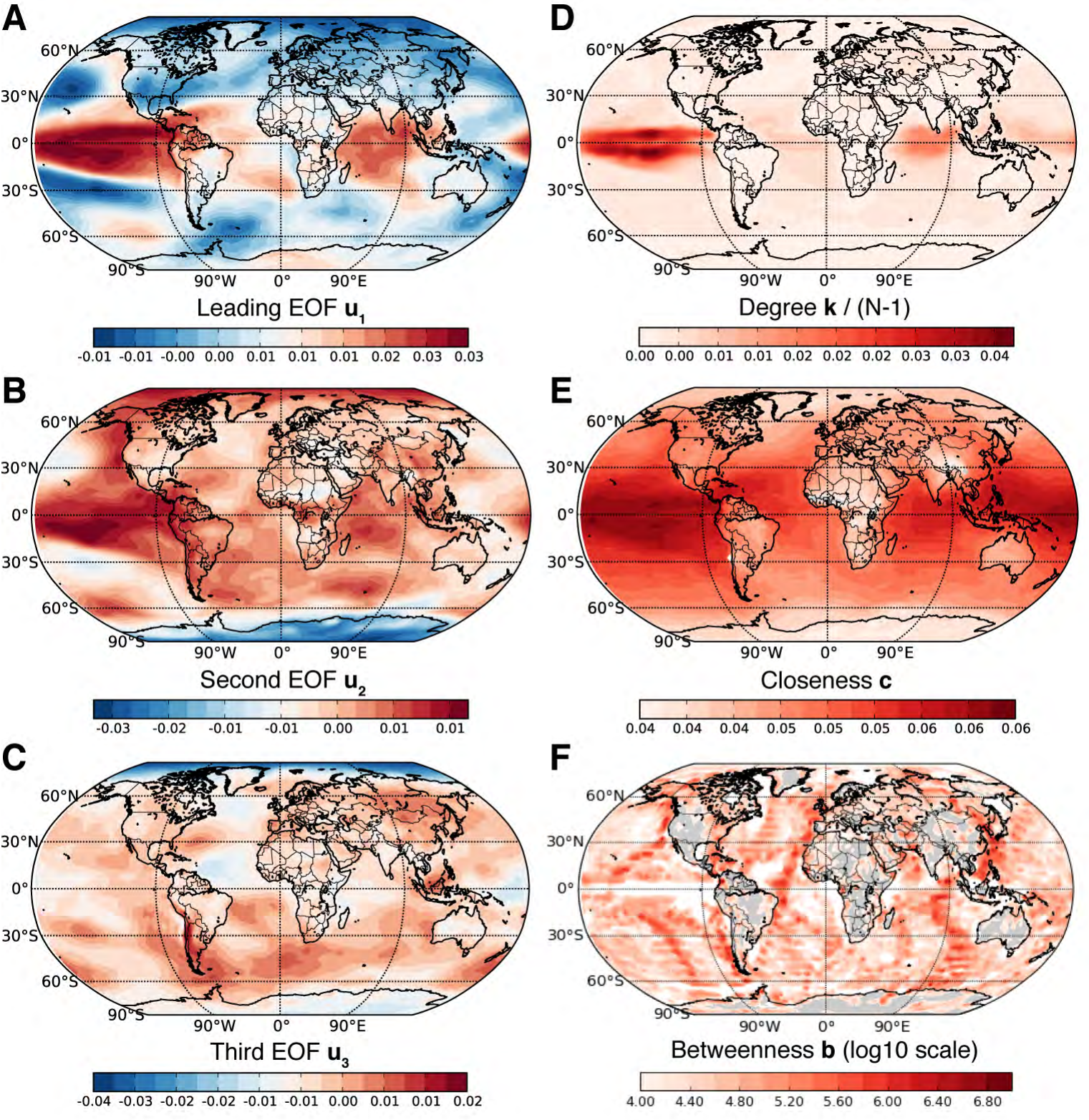}
\caption{Maps of (A,B,C) the leading three EOFs $\mathbf{u}_1$, $\mathbf{u}_2$, $\mathbf{u}_3$, (D) normalized degree field $\mathbf{k} / (N-1)$, (E)~closeness field $\mathbf{c}$, and (F)~betweenness field $\mathbf{b}$ for the global NCEP/NCAR surface air temperature climate network. The network construction threshold $T=0.67$ was chosen to yield a link density of $\rho=0.01$. In panel (F), gray shading indicates regions with betweenness values smaller than $10^4$.}
\label{fig:eof_network_comp_sat}
\end{center}
\end{figure*}

Next, we investigate the NCEP/NCAR reanalysis I surface air temperature (SAT) field as another frequently studied data set. The properties of this data are complementary to those of the precipitation field discussed above in two aspects:  (i) for the SAT data, the leading two EOFs explain approximately the same amount of variance (Fig.~\ref{fig:eigen_spectrum_sat}), while the leading eigenvalue separates more markedly from the remainder of the spectrum in the case of the precipitation data (Fig.~\ref{fig:eigen_spectrum}), and, (ii) the SAT field is known to display a stronger degree of spatial coherency than the precipitation field. In the light of the discussion in Section~\ref{sec:disc_precip_evap}, these two properties are reflected when comparing the leading three EOFs and network properties for the SAT data set (Fig.~\ref{fig:eof_network_comp_sat}). Firstly, the degree field resembles the leading EOF less than in case of precipitation data (Fig.~\ref{fig:eof_network_comp_sat}A,D), which is expected due to the weaker separation of the leading eigenvalues (Section~\ref{sec:relations_single_field} and Eq.~\ref{eq:degree_approx}). Consistently, the degree field displays an even less pronounced similarity to the second and third EOFs (Fig.~\ref{fig:eof_network_comp_sat}B,C,D). While the patterns found in the CC field (Fig.~\ref{fig:eof_network_comp_sat}E) still partly resembles those in the degree field (Fig.~\ref{fig:eof_network_comp_sat}D) as well as those in the leading two EOFs (Fig.~\ref{fig:eof_network_comp_sat}A,B), the BC field displays markedly distinct features (Fig.~\ref{fig:eof_network_comp_sat}F). Only in a few regions, these structures of high betweenness appear to coincide with patterns of large EOF loadings, e.g., high betweenness structures found along the West coasts of North and South America correspond to large positive loadings in the second and third EOFs, respectively.

The observed linear wave-like structures of large BC in the SAT field have been interpreted as signatures of the transport of temperature anomalies in strong surface ocean currents~\citep{Donges2009b,Donges2009a}. For example, the large betweenness structures resemble strong western boundary currents such as the Kuroshio of the east coast of Japan or Eastern boundary currents such the Canary current off the African west coast. It should be noted that while some of the structures in the BC field such as the one resembling the North Atlantic's subtropical gyre appear blurred, the logarithmic color scale in Fig.~\ref{fig:eof_network_comp_sat}F implies that even small changes in color correspond to exponentially large changes in BC. This interpretation of high betweenness structures in CNs constructed based on Pearson correlation as advective structures such as strong currents is supported by recent analytical studies that are based on well-known fluid dynamical model systems~\citep{Molkenthin2014a,Molkenthin2014b}. Further evidence that is also consistent with this interpretation of betweenness was found in a study of vertical interactions in the atmospheric geopotential height field, where regions of large cross-BC in the Arctic suggest that vertical air induced by the Arctic vortex is important for mediating the propagation of wind field anomalies between different isobaric surfaces \citep{Donges2011a}. Also, \citet{boers2013complex} employ BC and further network measures for precipitation data over South America to highlight the importance of atmospheric structures such as the South American low level jet for the propagation of extreme rainfall events, specifically over long distances.

\subsection{Potentials of climate network analysis} \label{sec:disc_potentials}

The examples discussed above suggest that CN analysis may be particularly useful in situations where (i) a dominant EOF (pair of coupled patterns) explaining significantly more variance (cross-covariance) in the data than further modes does not exist and (ii) the climatological field of interest displays a certain degree of spatial coherence reflecting, e.g., winds, ocean currents or long-range teleconnections. Such rules could be useful in practice when deciding on which methodology should be applied to a data set of interest. While future research beyond the scope of this work is needed to address these suggestions, we move on to discuss the potentials of CN analysis from a methodological point of view.

Considering higher-order network properties, approximate and exact relationships akin to Eqs.~(\ref{eq:degree_approx}) and (\ref{eq:degree_exact}) can be derived for other (coupled) CN measures of interest like the local clustering coefficient \citep{Donges2009a, Malik2012}

\begin{align}
\mathcal{C}_i &= \frac{\sum_{j,k=1}^N A_{ij} A_{jk} A_{ki}}{\sum_{j,k=1}^N A_{ij} A_{ik}} \label{eq:clustering}
\end{align}
by plugging in the approximation $A_{ij} \approx \Theta(|\lambda_1 u_{i1} u_{j1}| - T) - \delta_{ij}$ or the full expansion of $A_{ij}$ in terms of EOFs (Section~\ref{sec:relations_single_field}). However, the resulting lengthy expressions, particularly for path-based network measures such as CC and BC~\citep{Heitzig2012}, hardly help to gain further understanding other than that both eigen and network approaches are based on the same underlying similarity matrix (Figs.~\ref{fig:univariate_scheme} and \ref{fig:multivariate_scheme}). In contrast, taking the local clustering coefficient as an example illustrates the added value of the complex network point of view: Eq. (\ref{eq:clustering}) can be easily understood as a local measure for transitivity in the correlation structure of a climatological field~\citep{Donges2009a,Donges2011a}, while the same measure viewed as some function of all EOFs $\mathbf{u}_k$ would be considered hard to interpret or meaningless in terms of eigenanalysis alone. In that sense, the network approach allows insights into the correlation structure of climatological fields that go beyond and complement those obtainable by EOF analysis.

It has been shown in earlier studies that the statistical information provided by CN analysis is valuable for complementing standard techniques of eigenanalysis for tasks like model tuning, model validation~\citep{Feldhoff2014}, model and model-data intercomparison~\citep{Petrova2012,Steinhaeuser2013,Fountalis2013,Feldhoff2014}, statistical forecasting \citep{Steinhaeuser2011b}, and explorative data analysis~\citep{Steinhaeuser2010,Steinhaeuser2011}. Furthermore, the network approach allows to employ advanced algorithms for pattern recognition~\citep{Kawale2013}, spatial coarse-graining~\citep{Fountalis2013} or community detection~\citep{Tsonis2010,Steinhaeuser2011b,Steinhaeuser2014}. Recently, a series of studies based on well-defined fluid-dynamical model systems has provided deeper insights into the structure of CNs, particularly into how the latter is related to the dynamics of the underlying physical system, as well as fostered the interpretation of CN measures \citep{Molkenthin2014a,Molkenthin2014b,Tupikina2014}.

A particular advantage of CN analysis is that statistical methods originating from information and dynamical systems theory such as transfer entropy~\citep{Runge2012,Runge2012b}, probabilistic graphical models~\citep{EbertUphoff2012a,EbertUphoff2012b}, or event synchronization~\citep{Malik2012} can be naturally used for network construction, and, hence, for identifying processes and patterns which are not accessible when studying linear correlation matrices alone. Applying these modern methods of time series analysis for network construction allows, among other applications, to study the synchronization of climatic extreme events~\citep{Malik2012,boers2013complex,boers2014complex} or to suppress the misleading effects of auto-dependencies in time series, common drivers and indirect couplings by reconstructing causal interactions (in the statistical sense of information theory) between climatic sub-processes \citep{EbertUphoff2012a,Runge2012,Runge2012b,Runge2014}. This in turn enables a more direct interpretation of the reconstructed network structures and resulting patterns in network structures in terms of climatic sub-processes and their interactions, avoiding the conceptual problems that arise in the interpretation of results from purely correlation-based techniques such as classical EOF or CP analysis / MCA~\citep{Dommenget2002,Jolliffe2003,Monahan2009}.

\section{Conclusions} \label{sec:conclusions}

In summary, the main aim of this article has been to put the recently developed CN approach into context with standard eigenanalysis of climatological data, since both classes of methods are usually based on the same set of statistical similarity matrices, \textit{i.e.}, the linear correlation and cross-correlation matrices at zero lag. We have derived formal relationships between empirical orthogonal functions or coupled patterns and frequently used first-order CN measures such as degree or cross-degree, respectively. These relations have been illustrated empirically using global satellite observations of precipitation and evaporation fields as well as reanalysis data for the global surface air temperature field. However, it has been shown that, and in which specific practical settings, higher-order CN measures such as closeness and betweenness may contain complementary statistical information with respect to classical eigenanalysis. We have argued that this information could be valuable for tasks such as model tuning, validation, and intercomparison as well as for improving statistical predictions of climate variability and explorative data analysis. Hence, by transferring insights and tools from complex network theory and complexity science to climate research, CNs meet the need for novel techniques of climate data analysis facing quickly increasing data volumes generated by growing observational networks and model intercomparison exercises like the coupled model intercomparison project (CMIP)~\citep{Taylor2012}. Furthermore, the application of advanced network-theoretical concepts and methods from fields like complexity science, information theory and machine learning promises novel and deep insights into Earth system dynamics, particularly considering the complex interactions of human societies with global climatic and biogeochemical processes.

\begin{acknowledgements}
This work has been financially supported by the Leibniz association (project ECONS), the German National Academic Foundation, the Potsdam Institute for Climate Impact Research, the Stordalen Foundation, BMBF (project GLUES), the Max Planck Society, and DFG grants KU34-1 and MA 4759/4-1. For climate network analysis, the software package \texttt{pyunicorn} was used that is available at \texttt{http://tocsy.pik-potsdam.de/\-pyunicorn.php} \citep{Donges2013}. We thank Reik V. Donner and Doerthe Handorf for discussions and comments on an earlier version of the manuscript.
\end{acknowledgements}


\end{document}